\documentclass[twocolumn, amsmath, amssymb, superscript address] {revtex4-2}
\usepackage{hyperref}
\hypersetup{colorlinks=true, citecolor=blue, urlcolor=blue, linkcolor=blue}

\usepackage{graphicx}
\usepackage{dcolumn}
\usepackage{color}

\usepackage{bm}
\usepackage{amsmath,amssymb,graphicx}
\usepackage{epsfig}
\usepackage{enumerate}
\usepackage{array}
\usepackage{multirow}
\usepackage{gensymb}
\usepackage{comment}
\usepackage{import}
\usepackage{pdfpages}
\makeatletter
\AtBeginDocument{\let\LS@rot\@undefined}
\makeatother

\begin {document}

\title{Low frequency resistance fluctuations in an ionic liquid gated channel probed by  cross-correlation noise spectroscopy }
 
\author{\rm Bikash C. Barik}
\thanks{Contributed equally}
\affiliation{Department of Physics, Indian Institute of Technology Bombay, Mumbai, 400076, India}

\author{\rm  Himadri Chakraborti}
\thanks{Contributed equally}
\affiliation{Department of Physics, Indian Institute of Technology Bombay, Mumbai, 400076, India}
\affiliation{SPEC, CEA, CNRS, Universite Paris-Saclay, CEA Saclay, 91191 Gif sur Yvette, Cedex France}

\author{\rm Aditya K. Jain}
\thanks{Contributed equally}
\affiliation{Department of Physics, Indian Institute of Technology Bombay, Mumbai, 400076, India}
\affiliation{Department of Physics, Royal Holloway, University of London, Egham, Surrey, TW20 0EX, UK}

\author{Buddhadeb Pal}
\affiliation{Department of Physics, Indian Institute of Technology Bombay, Mumbai, 400076, India}
\affiliation{Indian Association for the Cultivation of Science, Kolkata, 700032, India}

\author{H.E. Beere}
\affiliation{Semiconductor Physics Group, Cavendish Laboratory, University of Cambridge, Cambridge CB3 0HE, U.K.}

\author{D.A. Ritchie}
\affiliation{Semiconductor Physics Group, Cavendish Laboratory, University of Cambridge, Cambridge CB3 0HE, U.K.}

\author{K. Das Gupta}
\email{kdasgupta@phy.iitb.ac.in}
\affiliation{Department of Physics, Indian Institute of Technology Bombay, Mumbai, 400076, India}
\begin{abstract}

 A  system in equilibrium keeps ``exploring" nearby states in the phase space and consequently, fluctuations can contain information, that the mean value does not. However, such measurements involve a fairly complex interplay of effects arising in the device and measurement electronics, that are non-trivial to disentangle.  In this paper, we briefly analyse some of these issues and show the relevance of a two-amplifier cross-correlation technique for semiconductors and thin films commonly encountered. We show that by using home-built amplifiers costing less than 10 USD/piece one can measure spectral densities as low as $\sim 10^{-18}-10^{-19}~ {\rm {V^2}{Hz^{-1}}}$. We apply this method to an ionic liquid gated Ga:ZnO channel and show that the glass transition of the ionic liquid brings about a change in the exponent of the low frequency resistance fluctuations. Our analysis suggests that a log-normal distribution of the Debye relaxation times of the fluctuations and an increased weight of the long timescale relaxations can give a semi-quantitative explanation of the observed change in the exponent.    
\end{abstract}
\maketitle

A key problem encountered in all measurements of ``noise'' is that the recorded data always contains the noise originating from the sample/DUT (Device under test) and the measurement circuit, in particular the first stage amplifier. There are two generic approaches to this aspect.

\begin{figure}[t]
	\centering
    \includegraphics[width=0.5\textwidth]{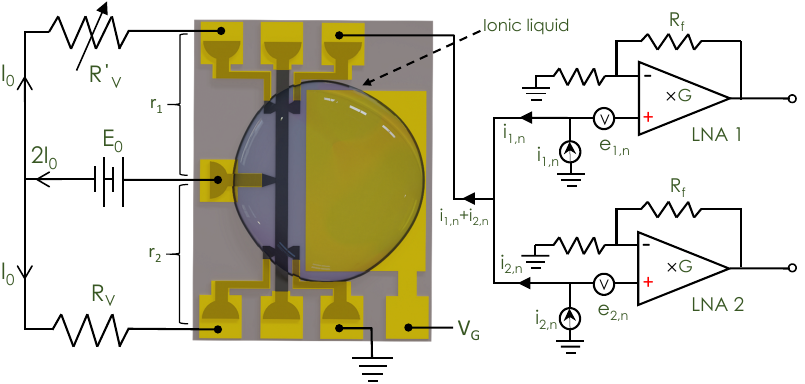}
    \caption{Circuit diagram of a five probe technique to measure resistance fluctuations through voltage fluctuations. Equivalent circuit for low noise amplifiers (LNA)  used in  cross-correlation measurements. The LNAs are powered by separate power sources. The dimension of the Hall bar is $\approx 1150 \times 100 {\rm \mu{m}}$. The typical value of $R_{V}\sim 250 \rm k\Omega$. The input voltage and current noise of the LNAs are shown as $e_{1,n}$, $e_{2,n}$, $i_{1,n}$ and $i_{2,n}$. The gain of each amplifier is $G$.} 
   \label{Five_point_bridge_schematic}
\end{figure}

\begin{enumerate}
    \item One ensures that the first stage amplifier's noise is assuredly less than the DUT. In literature \cite{scofield1987ac}
 it is often addressed by choosing a {\em passive} small-signal transformer (PST) as the first stage. These transformers with very high core susceptibility \cite{SR554, SR1900} often have noise contribution as low as $0.1 {\rm nV/\sqrt{Hz}}$ and certain more specialised designs can even be operated at cryogenic temperatures.  While these give excellent Common-mode-rejection (CMRR), the limitation is that the input impedances these can handle are limited (typically $\ll 1 {\rm k\Omega}$). It is important to point out that the impedance seen by the amplifier is always the ``2-terminal'' resistance which includes the typical lead-resistance of a cryostat (or similar equipment) and ohmic-contact resistance of the DUT itself. Even in cases where the 4-terminal resistance of the DUT is very small the resistance seen by the transformer is not necessarily small due to these inevitable contributions appearing in series. Another aspect is that since no transformer can be truly d.c. coupled, the very low frequency resistance fluctuations can only be measured by modulating them with a relatively higher frequency current, effectively translating the resulting voltage fluctuations to the most sensitive area of the transformer's own noise figure.  
    \item Another approach is to use reasonably low noise but {\em d.c. coupled} and relatively inexpensive (in our experiments, the chips used are all priced below 10 USD) BJT/FET input operational amplifiers, that can handle significantly higher input impedances but will have an intrinsic input noise typically a factor of $\sim 10-20$ higher compared to the PST. The resistances encountered in measurements on semiconductors \cite{peransin19901,lee2009low,shao2009flicker,watanabe2018remarkably,ma2019low}, carbon nanotube \cite{collins20001,appenzeller20071}, Indium arsenide nanowires \cite{delker20121, delker2013low} and 2D materials \cite{mehra2023origin,pal2009ultralow,pal2011microscopic,pal2009resistance,kamada2023suppression,karnatak2016current,kamada2021electrical,pal2010large,heller2010charge,moulick2022sensing} are often in the range of $1$-$100$ ${\rm k\Omega}$. In this method, the voltage signal is fed in parallel to two nearly identical first-stage amplifiers powered from independent and uncorrelated sources. Their outputs are then simultaneously digitised (in our experiments with $24$ bit resolution in $\pm 300~{\rm mV}$ range and at a rate of up to $200~{\rm kS/sec}$), discrete Fourier transformed (DFT)  after passing through an anti-aliasing filter and then {\em cross-correlated} in the frequency domain. In certain niche applications, like detecting shot noise \cite{oliver1999hanbury, kapfer2019josephson} or thermal noise \cite{dutta2022isolated, le2022heat}  in mesoscopic conductors for probing quantum statistics \cite{kapfer2019josephson, bartolomei2020fractional, glidic2023cross}, the granularity of charge \cite{assouline2023emission} etc., the signals are handled through similar techniques of {\em auto \rm{or} \em {cross}-correlation}. This is achieved by employing an initial-stage cryogenic amplifier (at $\sim 4$ K) with a typical gain of $5-7$  \cite{dicarlo2006system} combined with a room-temperature one. They are often designed to operate at a few MHz. This approach ensures the primary amplification of the signal in close proximity to the sample, preventing the accumulation of undesired signals before being seen by the room temperature electronics. As described \cite{jain2018cross, Supplementary_informations}  the cross-correlation process largely (but not entirely) eliminates the uncorrelated noise from the amplifiers themselves. The resulting output is the spectral density of the signal that was ``common'' to both the amplifiers. While this suffices for a large class of experiments with Gaussian fluctuations, it should be borne in mind that the original time series cannot be reconstructed.
    
    \end{enumerate}

 \begin{figure}[b]
	\centering
    \includegraphics[width=0.47\textwidth]{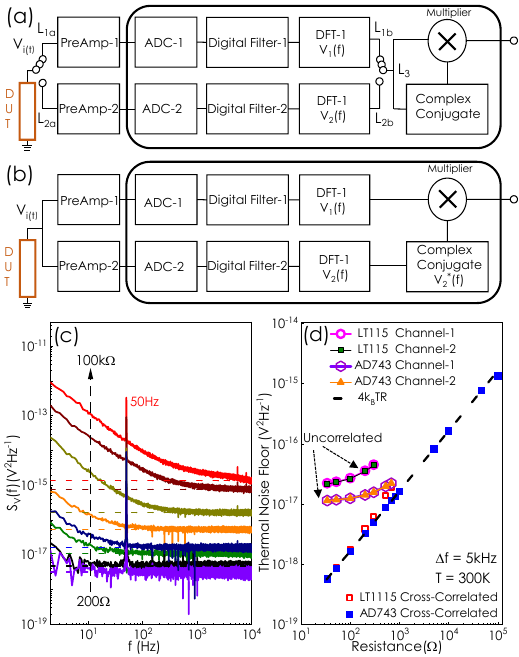}
    \caption{(a) Shows schematic for noise measurement using a single channel with a pre-amplification stage, contains excess noise introduced by the preamplifier and other electronic components, (b) shows schematic for noise measurements using cross-correlation of two channels with two independently powered amplifier, cancelling out the excess noise from the electronic components. (c) shows PSD recorded for various DUTs from $200\ohm$ to $100 \rm k\ohm$ (solid lines) and corresponding average noise floor (dashed line). (d) The magnitude of the noise floor recorded for different DUTs from a single channel and two-channel cross-correlation are shown for two different LNAs. The black dashed line shows the expected thermal noise $4k_{B}TR$. }
   \label{standard_resistors}

\end{figure}

 Our implementation of the cross correlation technique is shown in  Fig. \ref{Five_point_bridge_schematic}. The sample itself forms {\em two} arms, assumed to have resistances $r_1$
 and $r_2$, of a Wheatstone bridge with much larger resistances $R_v$ and $R'_{v}$ forming two other arms. One of the large resistances is adjustable and is used to null the bridge output. We have $r_1\approx{r_2}$ and $R_{v}, R_{v}^{'} \gg r_1,r_2 $. The output is then largely immune to small fluctuations of the biasing source ($E_0$) and the ambient temperature, as fluctuations arising out of these will tend to affect $r_1$ and $r_2$ almost identically. However, a small resistance fluctuation in one of the arms (due to internal processes like trapping-de-trapping, thermal generation etc.) will be picked up. This signal is then fed to both the amplifiers (each of gain $G$), as shown and their outputs cross-correlated. The expected (cross-correlated) power spectral density of the voltage fluctuations arising from these resistance  fluctuations, when current $I_0$ flows through the DUT is given by Eq. (\ref{PSD_relation_with_current_flow})\cite{scofield1987ac},

\begin{equation}
    S_{v}(f,I_0) = {G^2}\left[ S_{v}^{0}(f) + {{I_{0}}^2}S^{-}_{r}(f)\right]
\label{PSD_relation_with_current_flow}
\end{equation}

where $S_{v}^{0}(f)$  is the power spectrum of the background originating from electronic and thermal noise, $S^{-}_{r}(f)$ is the spectrum of ${\delta}r_{-}(t) = {\delta}r_{1}(t) - {\delta}r_{2}(t)$ with long time average $\left\langle {\delta}r_{1}(t){\delta}r_{2}(t)\right\rangle = 0$.   The choice of the first-stage amplifier is guided by a balance between the inevitable intrinsic input current and the current noise of the operational amplifier itself. Fig. \ref{Five_point_bridge_schematic} shows the equivalent circuit with the noise equivalent sources included as part of the input stages. It is not only the DUT voltage but the combination (see Fig. \ref{Five_point_bridge_schematic}) of three voltages that would get amplified. They are related as follows :
 
 \begin{equation}
     v_{i,\rm{rms}}^2  = v_{\rm DUT}^2 + e_{\rm i,n}^2 + \left( i_{1,\rm{n}} + i_{2,\rm{n}}\right)^{2}{R_{\rm DUT}}^2 \ldots i=1,2
     \label{rms_addition_of_current_and_voltage_noises}
 \end{equation}
 
 Ultimately this determines the practicality of the measurement. We find that a current noise not exceeding few {$ fA/\sqrt{\rm Hz}$} and a voltage noise not exceeding $\sim 5 {\rm nV/\sqrt{Hz}}$ are essential for most measurements. Choosing an amplifier with lower voltage noise but higher current noise is detrimental once the resistance of the DUT becomes slightly higher, in which case the noise current flowing through the DUT manifests itself as a much higher voltage noise over-ruling the benefit of a lower voltage noise. Although, for very low source resistances, this would be acceptable. However, our purpose is to design an inexpensive and generic setup covering the typical range of resistances encountered in semiconductors and similar samples whose resistances may well go up to several tens of $\rm{k\Omega}$. This is illustrated with two commonly available operational amplifiers in Fig. \ref{standard_resistors}(d). If no cross-correlation is used, the spectrum obtained corresponds to the two upper curves (Uncorrelated). However, once the two data are cross-correlated the contribution from the amplifiers is suppressed to a large extent and the correct thermal noise (with no adjustable parameters) is obtained over more than four orders of magnitude, spanning $ 10\Omega$ - $100 {\rm k\Omega}$. 
 
\begin{figure}[t]
	\centering   \includegraphics[width=0.48\textwidth]{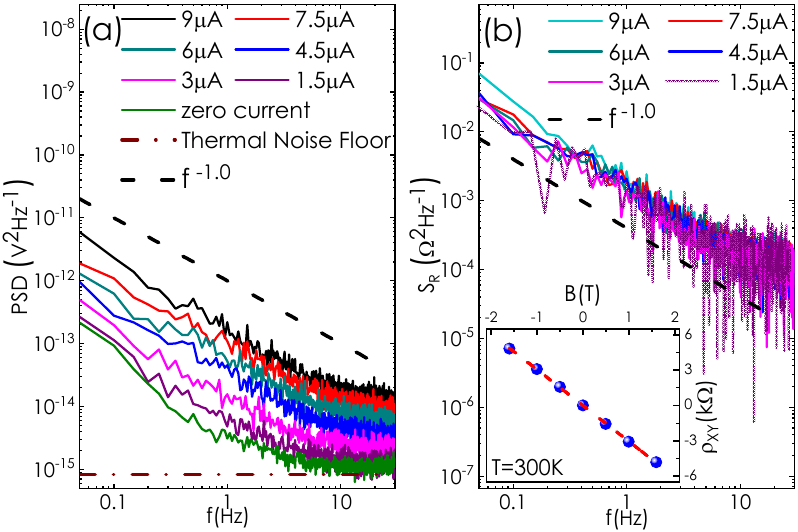}
    \caption{(a) Spectrum recorded at lower frequencies for GaAs/AlGaAs heterostructure for different current values at room temperature, (b) Resistive noise obtained from the spectra after scaling. The scaling is done by subtracting the zero current data from each trace and dividing by $I^2$ following Eq. (\ref{PSD_relation_with_current_flow}). Hall data at room temperature is shown in the inset. The slope, $\rm 3.045 \ k\ohm T^{-1}$, corresponds to a two-dimensional carrier density of $2.07\times{10^{11}}~{\rm cm^{-2}}$.}
   \label{GaAs/AlGaAs_noise}
\end{figure}

The typical behaviour of $S_{v}(f)$ has a well defined nearly flat high frequency ($f \geq 100 {\rm Hz}$) part and a low frequency ($0.01 \leq f \leq 10 {\rm Hz}$) $\propto \dfrac{1}{f^{\gamma}}$, with $\gamma\approx{1}$ (Fig. \ref{standard_resistors}(c)). The correctness of the absolute values produced by the cross-correlation technique can be judged by two observations.
\begin{enumerate}
    \item The $f \gtrsim 100~{\rm Hz} $ part of our measurements agrees very closely with $S_{v}(f)=4{k_B}TR$, where $R$ is the 2-probe resistance ``seen'' by the amplifier (Fig. \ref{standard_resistors}(c) \& (d)) is the well known Johnson-Nyquist noise \cite{nyquist1928thermal,reif1998fundamentals}
    \item The correctness of the low-frequency part can be established by ascertaining the magnitude of the Hooge parameter \cite{hooge1981experimental, dutta1981low} $\alpha_H$ given by
    
    \begin{equation}
       S_{r}(f) = \alpha_{H}  \frac{r^2}{Nf^{\gamma} }
       \label{Hooge_paramter_definition}
    \end{equation}
    
     Notice that this requires knowing the {\em total} number of carriers ($N$) participating in the conduction process. We used a well-characterised 2D electron gas (2DEG) in a GaAs-AlGaAs High electron mobility transistor (HEMT) $100 {\rm nm}$ below the surface and hence mostly free from the effect of scattering from the charged surface states and oxide monolayers. Hall measurements on the HEMT yielded a carrier density of $2.07\times{10^{11}}~{\rm cm^{-2}}$ and a room temperature mobility of $7650~{\rm cm^{2}.V^{-1}s^{-1}}$  The dimensions of the etched mesa was $1250 \times 140~ {\rm \mu{m}}$ and the resistance between the pair of contacts used was $r~=~ 20.5 {\rm k\Omega}$ at $T\approx300~{\rm K}$. Using the data of Fig. \ref{GaAs/AlGaAs_noise}(b) and the above-mentioned numbers we get $\alpha_{H} \approx  {1.4}\times{10^{-3}}$, which is very close to the values reported by Hooge \cite{dutta1981low, hooge1981experimental} when the origin of the noise is primarily in the bulk.
\end{enumerate}

\begin{figure*}[t]
	\centering
    \includegraphics[width=0.85\textwidth]{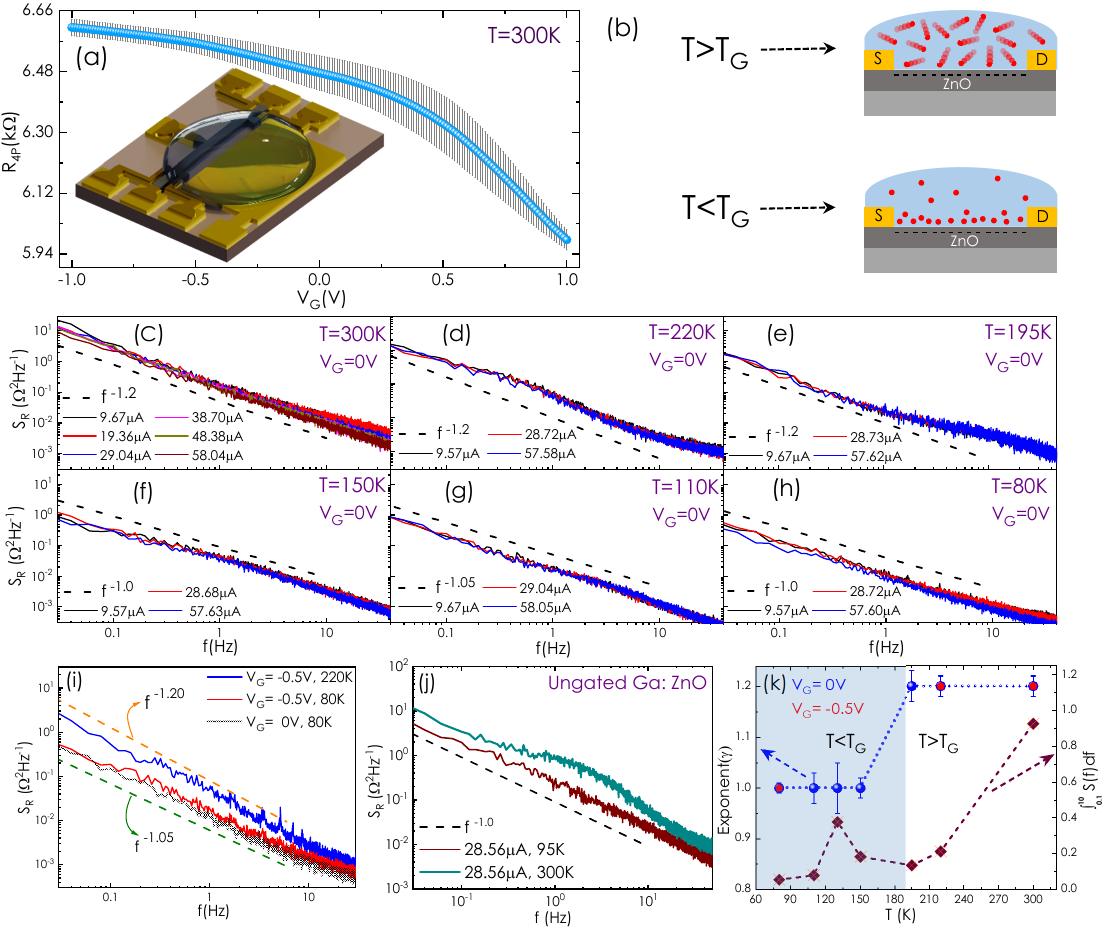}
    \caption{ (a) At T=300K, the transfer characteristics exhibit a nearly hysteresis-free nature of the IL-gated device. (The average of multiple curves with standard deviation is shown as a grey band). A schematic of the device is shown in the inset. (b) The motion of mobile ions shown, which are frozen below the glass transition temperature ($ T<T_{G}$). (c)-(h) The observed normalized low-frequency noise at temperatures of 300K, 220K, 195K, 150K, 110K, and 80K respectively. Currents applied are labelled. In all the above cases (c-h), the gate is always grounded. (i) Normalized noise spectra at 220K and 80K for applied $V_G$= -0.5V. $V_{G}$=0V at 80K is shown for reference. (j) Normalized noise spectra for ungated ZnO thin films show $1/f$ noise at $\rm 95K$. (k) Summarizes the behaviour of the exponent and the integrated noise power over the entire temperature range.}
   \label{IL_gated_ZnO_noise}
\end{figure*}

The presence of a parallel layer of charge at some distance from a 2D electron or hole gas (2DEG/2DHG) is a common feature in heterostructures. Either the modulation doped layer, a delta doped layer or the charged surface states (in the case of a shallow 2DEG) create an impurity potential in the plane of the 2D system. This potential, screened by the 2DEG and suitably averaged over its form factor is one of the major sources of scattering that limits the transport \cite{Ando_Fowler_Stern_RMP, DaviesSemiconductorBook, WendyMak_shallowAPL, KozolovSemiconductors_2007} as well as the quantum lifetime \cite{DQWang_PRB_2013, hong2009quantum}. While for relatively lower carrier density systems like MOSFETs, GaAs-AlGAs HEMT and HHMTs (High hole mobility transistors) a dielectric-metal gate is used to modulate the carrier densities,  the technique becomes untenable when the carrier densities are very high. In several systems of recent interest \cite{daghero2012large,sagmeister2006electrically,petach2014mechanism,piatti2017control,piatti2022reversible,yuan2009high} modulation of carrier densities is possible only via ionic liquid (IL) gates capable of forming self-organized nano gap capacitors upon application of gate voltage, resulting in capacitance in the range of $\sim 10-50 \mu F cm^{-2}$ \cite{lu2015evidence, costanzo2016gate, daghero2012large, petach2014mechanism, ueno2011discovery, piatti2017control, yuan2010electrostatic}. For the IL gated system, the electric double layer is effectively a very high density surface charge. However, systems gated with IL are also inherently very high carrier density systems and the impurity Coulomb potential would be screened. Thomas-Fermi screening gives $V(q) \sim \dfrac{1}{q^2 + k_{TF}^2}$ where the Thomas-Fermi wave vector is given by $k_{TF} = \left( \dfrac{{e^2}D(E_F)}{{\epsilon_r}{\epsilon_0}}\right)^{1/2}$. The density of states at the Fermi level $D(E_F)$ makes $k_{TF}$ sufficiently large to suppress large angle scattering which has a maximal effect on transport lifetime ({\em i.e.} mobility ). However, it is known that as the IL nears the glass transition temperature $T_G$ from the high temperature side, the short time fluctuations are gradually frozen out irrespective of the gate voltage. It is seen experimentally that most of the time the resistivity of the conducting channel itself does not have any sharp feature as the IL crosses $T_G$ (for DEME-TFSI used in our experiments $T_G \approx 180-200 {\rm K}$). However, our investigations show that the exponent $\gamma $ of the low frequency resistance noise ( $S_V(f)\propto 1/f^{\gamma}$,  using the method described earlier) in an IL gated ${\rm ZnO}$ has a pronounced change. The results are valid over three to four decades of frequency.  While transport and photo-response using an ionic liquid ($\rm PEO+LiClO_{4}$) gate has been reported n-ZnO samples \cite{ghimire2015synergistic,mondal2015mobility}, the temperature dependence of low frequency noise in these IL-FET has not been previously explored.     

Our sample (Fig. \ref{IL_gated_ZnO_noise}(a)) is a Ga-doped ZnO thin film deposited on c-sapphire by RF magnetron reactive co-sputtering of Zn and $\rm GaAs$ at 375\degree C with 3\% of $\rm GaAs$ coverage on the Zn target, in $\rm Ar-O_{2}$ sputtering atmosphere. Subsequently, these are lithographically patterned into Hall bars with $W=$100 $\mu m$ and $L$=1150 $\mu m$ and $\rm Ti/Au$ contact pads.  
Hall measurement \textcolor{blue}{(Supplementary data)} gives a carrier density of \ $ \rm 9.97 \times 10^{19} \ {cm}^{-3}$. The four-terminal resistivity is $\rho
 = \left(\frac{W}{L}\right) \times R_{xx} = 4.22 \times 10^{-3} \ \Omega \ \text{cm}$. This  yields a  Hall mobility  $\mu_H = \frac{1}{\rho n_{3D} e} = 14.85 \ \text{cm}^2 \ \text{V}^{-1} \ \text{s}^{-1}$, at $\rm 300K$.  
While applying on the channel, the contacts were covered with silicone gel to protect them \cite{misra2007electric}.  After application, it was cured in a high vacuum at $\rm 110 \degree \ C$  for $\sim$ 48 hrs. By making the gate electrode much larger than the channel area,  the necessity of a reference electrode was eliminated since most of the potential will drop across the working electrode \cite{yuan2010electrostatic}. At room temperature, the gate voltage sweep shows a near $10\%$ variation of the resistance for applied $V_G$ of $\pm 1\ {\rm V}$, shown in Fig. \ref{IL_gated_ZnO_noise}(a).  

\begin{figure}[t]
	\centering
    \includegraphics[width=0.47\textwidth]{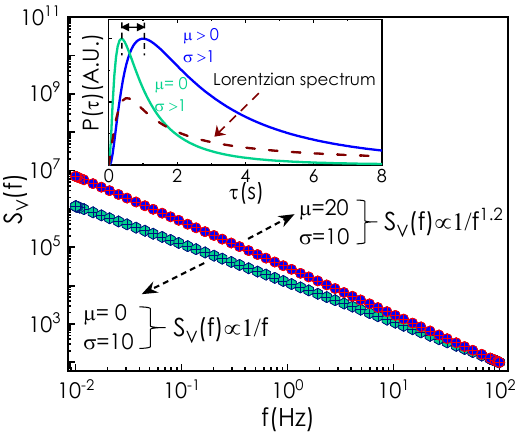}
    \caption{The numerically computed $S_{V}(f)$ using Eq. (\ref{log_normal_distribution}), successfully replicates the variation of the exponent $\gamma$ observed in the experiment. (inset) The Lorentzian spectrum (dotted line) and the log normal distribution function (solid lines) are depicted schematically. The shift of the peak of the log-normal ultimately controls the power law.}
   \label{ Log_normal_distribution_function_calculation}
\end{figure}

The measurement of noise was carried out from $T\approx 300 {\rm K}$ to well below $T_G$, as shown in Fig. \ref{IL_gated_ZnO_noise}(c)-(j). We find that for $T > T_G$ the exponent $\gamma \approx 1.2$ with no strong dependence on the gate voltage. For $T \ll T_G$, it is clear that $\gamma\approx 1$, as summarised in Fig. \ref{IL_gated_ZnO_noise}(k). The magnitude of the noise has a small peak around the transition region. 
The $1/f$ noise at low frequencies is thought to arise from fluctuations with Debye-type relaxation having distributions of the relaxation time $\tau$ over several decades of frequencies. The distribution of the relaxation times $P(\tau)$ indeed affects the power law \cite{hooge1981experimental}.  We show that by assuming a physically reasonable log-normal distribution whose mean ($\mu$) is slightly shifted but the spread ($\sigma$) is essentially kept constant, we can reproduce in a simple parametric way, variations of the exponent of the low frequency fluctuations. This can be expressed as, 

\begin{eqnarray}
\nonumber
S_{v}(f) &
=& \int_{0}^{\infty} {d\tau}P(\tau)\frac{\tau}{1 + \left( 2\pi{f
}\tau\right)^2} \\
P(\tau) &\propto& \frac{1}{\tau} \exp \left( -\frac{(\ln \tau - \mu)^2}{2\sigma^2} \right)
\label{log_normal_distribution}
\end{eqnarray}
where, $P(\tau)$ is the log normal distribution function. 
Fig. \ref{ Log_normal_distribution_function_calculation} shows the results of the numerically computed integral from Eq. (\ref{log_normal_distribution}) with the mean of the distribution shifted slightly towards the higher values of $\tau$ (as shown in the inset). The physical interpretation of this is that the weightage of the faster fluctuations, {\em i.e.} smaller values of $\tau$ are slightly reduced. This is consistent with the expectation that as the temperature is reduced below the glass transition temperature $T_{G}$ some of the relatively faster fluctuations are essentially frozen, affecting the distribution over which the Lorentzian relaxations must be summed over. 

In conclusion, we have presented a relatively inexpensive method of measuring low frequency noise that largely eliminates the noise originating in the measurement electronics. The method is also suited for a very wide range of input impedances that are encountered in many systems of current interest. We applied this method to an IL-gated degenerate semiconductor and showed that the exponent of the $1/f$ noise changes as the ionic liquid undergoes the glass transition. We linked this to a physically reasonable distribution of the relaxation times. While we have not tried to do a microscopic first principle derivation of this distribution, we note that it is a reasonable and simply parametrizable distribution that appears to correctly reproduce the nature of variation of the exponent of the $1/f$
 noise observed in our system as well as several others. The distribution can be further tuned in two significant ways that may be of physical importance. First, the spread of the distribution itself may be related to certain aspects of the interfacial defects. In addition, if there are specific recombination-generation processes occurring with a certain time scale \cite{ghosh2024electronic} an additional weight at that time scale can be added to the smooth background to reproduce the humps that are sometimes observed in this type of measurement. \\

Acknowledgements: BCB and AKJ acknowledge financial support through DST-INSPIRE and UGC fellowships of Govt. of India. Low temperature measurements were supported by the cryogenic facilities of IIT Bombay.  We thank S.K. Appani and S.S. Major (Dept of Physics, IIT Bombay) for providing the Ga:ZnO samples.

\bibliographystyle{apsrev4-2}
\bibliography{Noise_ionic_liquid_V1}

\onecolumngrid
\clearpage
\mbox{}


\section*{Supplementary material}

# Supplementary data: Low frequency resistance fluctuations in an ionic liquid gated channel probed by cross-correlation noise spectroscopy

Bikash C. Barik,[1, *] Himadri Chakraborti,[1, 2, *] Aditya K. Jain,[1, 3, *]
Buddhadeb Pal,[1, 4] H.E. Beere,[5] D.A. Ritchie,[5] and K. Das Gupta[1, †]

[1]*Department of Physics, Indian Institute of Technology Bombay, Mumbai, 400076, India*
[2]*SPEC, CEA, CNRS, Universite Paris-Saclay, CEA Saclay, 91191 Gif sur Yvette Cedex France*
[3]*Department of Physics, Royal Holloway, University of London, Egham, Surrey, TW20 0EX, UK*
[4]*Indian Association for the Cultivation of Science, Kolkata, 700032, India*
[5]*Semiconductor Physics Group, Cavendish Laboratory,*
*University of Cambridge, Cambridge CB3 0HE, U.K.*

## I. THE CROSS CORRELATION MEASUREMENT TECHNIQUE

Consider a discrete sequence of length $N$, given by $v[t]$ where $t \dots 0, 1, \dots N-1$. The Discrete Fourier Transform (DFT) of this sequence is given by

$$\tilde{v}[f] = \sum_{t=0}^{N-1} v[t] e^{-\frac{2\pi i t f}{N}} \tag{1}$$

and its inverse

$$v[t] = \frac{1}{N} \sum_{f=0}^{N-1} \tilde{v}[f] e^{\frac{2\pi i f t}{N}} \tag{2}$$

The power spectral density of the signal is given in terms of its Fourier components as

$$S_v[f] = \frac{1}{N} \tilde{v}[f] * \tilde{v}^*[f] \tag{3}$$

Now consider the time domain signal $v[t]$ entering two amplifiers each with gain $G$, but which also adds uncorrelated noises $n_1[t]$ and $n_2[t]$ to its ideal output $Gv[t]$. The outputs of the two amplifiers are denoted by uppercase $V$ and their DFT are given by

$$\tilde{V}_1[f] = G\tilde{v}[f] + \tilde{n_1}[f] \tag{4}$$
$$\tilde{V}_2[f] = G\tilde{v}[f] + \tilde{n_2}[f]$$

Then the cross correlated quantity is averaged over several traces (typically a few hundred in our experiments)

$$\frac{1}{N} \tilde{V}_1[f] \tilde{V_2}^*[f] = \frac{1}{N} \left( G\tilde{v}[f] + \tilde{n_1}[f] \right) \left( G\tilde{v}[f] + \tilde{n_2}[f] \right) \tag{5}$$

$$= \frac{1}{N} \left( G^2 \tilde{v}[f] \tilde{v}^*[f] + G\tilde{v}[f] \tilde{n_2}^*[f] + G\tilde{v}^*[f] \tilde{n_1}[f] + \tilde{n_1}[f] \tilde{n_2}^*[f] \right) \tag{6}$$

$$\left\langle \frac{1}{N} \tilde{V}_1[f] \tilde{V_2}^*[f] \right\rangle = \frac{G^2}{N} \left\langle \tilde{v}[f] \tilde{v}^*[f] \right\rangle + \frac{G}{N} \left\langle \tilde{v}[f] \tilde{n_2}^*[f] \right\rangle + \frac{G}{N} \left\langle \tilde{v}^*[f] \tilde{n_1}[f] \right\rangle + \frac{1}{N} \left\langle \tilde{n_1}[f] \tilde{n_2}^*[f] \right\rangle \tag{7}$$

$$\left\langle \frac{1}{N} \tilde{V}_1[f] \tilde{V_2}^*[f] \right\rangle \approx \frac{G^2}{N} \left\langle \tilde{v}[f] \tilde{v}^*[f] \right\rangle = G^2 S_v[f] \tag{8}$$

This happens because except for the first term in RHS of Eq. (7) all the terms contain long term averages of uncorrelated quantities which tend towards zero, which forms the basis of the technique. It is important to ensure that $N_1[t]$ and $N_2[t]$ are as uncorrelated as possible, for example by running the two amplifiers from independent power sources and in separate shielded boxes.

---

* Contributed equally
† kdasgupta@phy.iitb.ac.in

## II. IMPLEMENTATION OF THE CROSS CORRELATION TECHNIQUE OF ELECTRONIC NOISE MEASUREMENT

In order to verify the cross correlation method and usefulness over conventional single channel measurements, fluctuations from commercially available carbon/metal film resistors are per-amplified by two distinct and independently powered low noise amplifiers (AD743/AD745/LT1115) by a gain of ($\approx 1000$) in non-inverting configuration. The output of the low noise amplifiers (LNAs) is then fed to the two distinct channels of a DAQ (24-bit resolution). A code is written to acquire the data from the two channels, digitize them, take the Fourier transform, conjugate one of the transforms, multiply to the other transform for each frequency component and then update the average spectrum after every measurement. A sufficiently long measurement (to average 50-100 streams) can provide the rectified correlated signal and cancel all practical purposes uncorrelated noise, introduced by the two preamplifiers. Indeed the sensitivity is limited in a real situation by the finite measuring time and the residual correlation of the noises between the two channels due to the limitations of the preamplifiers and analog to digital converters.

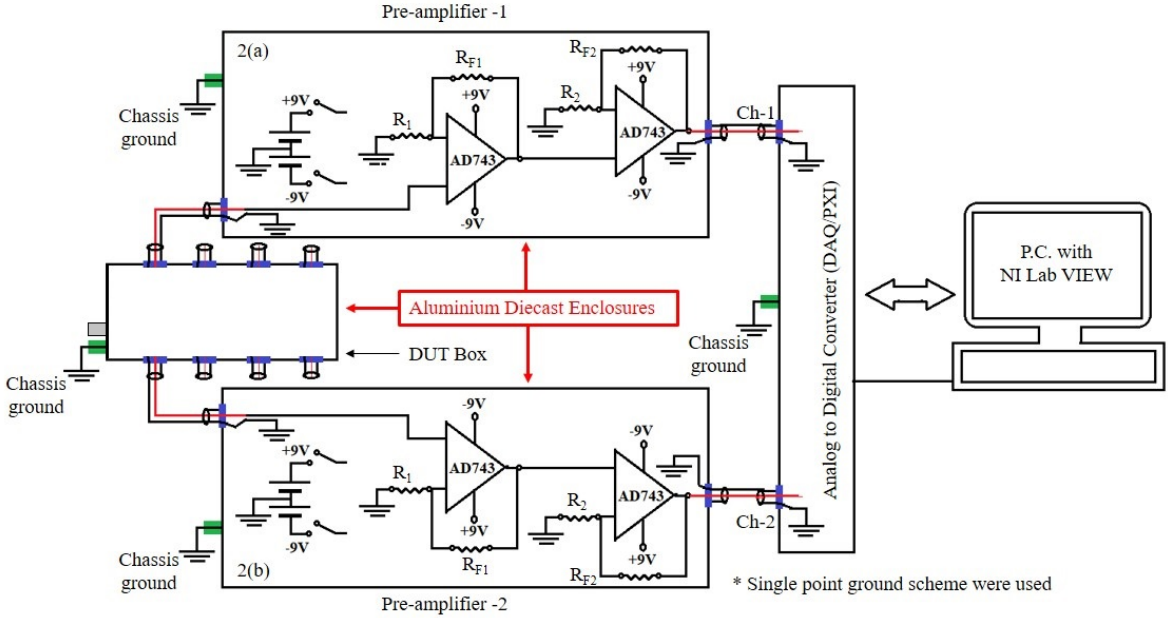

FIG. S1. Implemation of the cross correlation method.

Noise associated with any practical operational amplifier can be represented by combination of voltage equivalent noise source $e_n$, and a current equivalent noise source $i_{n\pm}$. Input voltage noise can always be represented by a voltage source in series with the non-inverting terminal of the operational amplifier. Input current noise can always be represented by current sources from both inputs to ground (Fig. S2). The resultant voltage noise $v_{rms}$ for a source resistance $R_s$ is given by Eq. (9).

$$v_{rms} = \sqrt{\left(v_{Nyquist}{}^2 + e_n{}^2 + \left\{(i_{n-} + i_{n+})R_s\right\}^2\right)} \tag{9}$$

The practical limitations of cross correlation measurement of electronic noise using low noise preamplifiers are limited by inherent noises associated with the preamplifiers. To understand that one can consider an operational amplifier with voltage noise $3nV/\sqrt{Hz}$ and current noise $1pA/\sqrt{Hz}$. With zero source impedance, the voltage noise $e_n$ will dominate since $v_{Nyquist}$ and $R_s$ both are zero. With a source resistance of $3k\Omega$, the current noise of $1pA/\sqrt{Hz}$ flowing in $3k\Omega$ will equal the voltage noise ($v = iR = 3nV/\sqrt{Hz}$), but the Johnson noise of $3k\Omega$ is $\sqrt{4k_BTR} = 7nV/\sqrt{Hz}$ and is dominant. With a source resistance of $300k\Omega$, the current noise portion increases $100\times$ to $300nV/\sqrt{Hz}$,

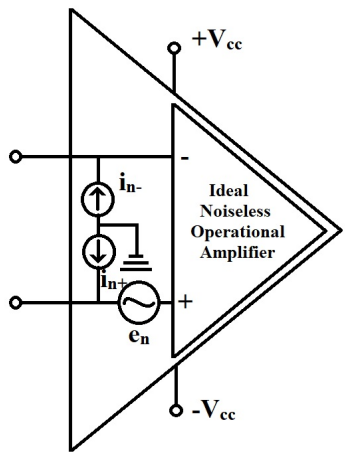

FIG. S2. Noise equivalent circuit of an operational amplifier.

voltage noise continues to be unchanged and the thermal noise increases 10 folds. Thus current noise dominates. This shows that the choice of a low noise operational amplifier depends on the source impedance of the signal and at high impedances, inherent current noise associated with the operational amplifiers always dominates. A bipolar field effect transistor amplifiers tend to have comparatively higher voltage noise but very low current noise. The recent amplifiers like AD743/AD745 have very low voltage as well as very low current noise. And can be used for voltage noise measurements for most of the practical purposes for which impedance is associated with the DUT up to $100k\Omega$.

### 1. Features of LT1115 Low Noise Amplifier

1. Voltage Noise $1.2\,\mathrm{nV}/\sqrt{\mathrm{Hz}}$ Maximum at $1\,\mathrm{kHz}$, and $0.9\,\mathrm{nV}/\sqrt{\mathrm{Hz}}$ typical at $1\,\mathrm{kHz}$.

2. Current Noise $2.2\,\mathrm{pA}/\sqrt{\mathrm{Hz}}$ Maximum at $1\,\mathrm{kHz}$, and $1.2\,\mathrm{pA}/\sqrt{\mathrm{Hz}}$ typical at $1\,\mathrm{kHz}$.

3. Gain-Bandwidth Product $40\,\mathrm{MHz}$ Minimum

4. Slew rate $10\,\mathrm{V}/\mu s$ Minimum

### 2. Features of AD743 Low Noise Amplifier

1. Voltage Noise $2.9\,\mathrm{nV}/\sqrt{\mathrm{Hz}}$ Maximum at $10\,\mathrm{kHz}$, and $0.38\,\mu\mathrm{V_{p-p}}$, $0.1\,\mathrm{Hz}$ to $10\,\mathrm{Hz}$.

2. Current Noise $6.9\,\mathrm{fA}/\sqrt{\mathrm{Hz}}$ Maximum at $1\,\mathrm{kHz}$.

3. Gain-Bandwidth Product $4.5\,\mathrm{MHz}$ Minimum.

4. Slew rate $2.8\,\mathrm{V}/\mu s$ Minimum.

## A. Design of the Preamplifier used for Noise Measurement

To perform noise measurements for large bands of frequencies, the gain bandwidth product of a low noise amplifier should be high. Therefore one has to design a preamplifier with high constant gain for a large bandwidth. An additional benefit of the large bandwidth is noise floor associated with the fluctuations can be measured. Measurement of voltage/current fluctuations is done in the frequency domain. Therefore it becomes important to design the electronics in such a way so that, gain should remain the same for all frequencies of practical interest. It is not very trivial to make such operational amplifiers, with ultra low current and voltage noise with very high bandwidth product. Therefore for different devices with different input resistances, one can choose different low noise amplifiers

for noise measurements accordingly.

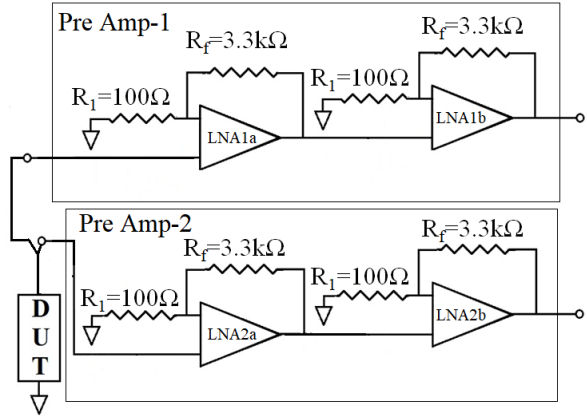

FIG. S3. Design of pre amplifiers using AD743 low noise amplifiers.

In our experiments we used AD743 low noise amplifier with a gain bandwidth product of 4.5 MHz, which gives us the freedom to choose the gain as high as $G = 1000$, with a spectrum of bandwidth not more than 2.25 kHz, with single stage amplification. The limit of bandwidth of noise measurement can be improved by adopting the double stage amplification. We designed low noise preamplifiers with a gain of $34 \times 34 = 1156$, using two stages of non-inverted amplification, each of non inverting mode Gain $= (1 + \frac{R_f}{R_1}) = 1 + 33 = 34$ (Fig.S3). Splitting of gain in two stages allows, one to choose a larger bandwidth for noise measurement compared to the single stage. By splitting of gain in two stages new range of bandwidth becomes 66 kHz, $\approx 30$ times higher than single stage amplification.

## III. TRANSPORT MEASUREMENTS

### A. Hall measurement on Zinc Oxide thin film

Fig. S4 presents the room temperature Hall measurement data of a ZnO thin film, indicating a two-dimensional carrier density of $7.48 \times 10^{14}$ cm$^{-2}$. The Hall data was recorded in the absence of an ionic liquid.

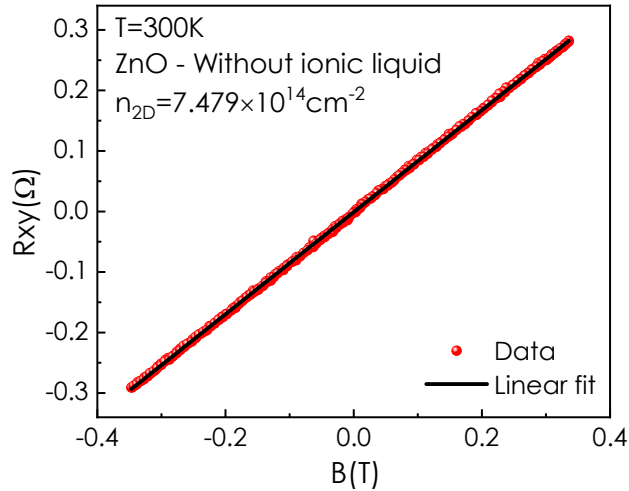

FIG. S4. Carrier density of ZnO thin film in the absence of ionic liquid.

## B.  Glass transition measurement of DEME-TFSI ionic liquid

Fig. S5 shows the behaviour of an ionic liquid gate. The ionic motion within the liquid starts to undergo a slowdown below approximately $\sim 230$K. Furthermore, below a specific temperature marked as the glass transition temperature ($T_G$), the ions associated with the ionic liquid experience complete immobilization. [1][2]. This happens over a range of $T \approx 180 - 200$K which corresponds well to the known value of $T_G$ of DEME-TFSI. Changing the gate voltage will have the desired effect only if done at $T \gg T_G$.

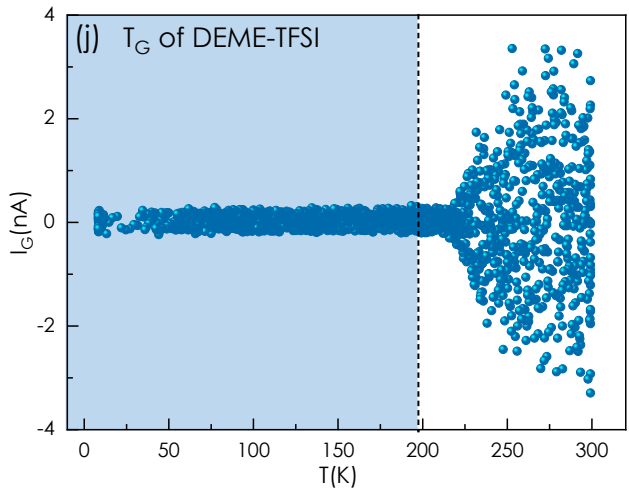

FIG. S5. The glass transition temperature of the ionic liquid demonstrates the freezing of ions below a certain temperature, as discussed in the main text.


\end{document}